# Unambiguous detection of mesospheric $CO_2$ clouds on Mars using 2.7 μm absorption band from the ACS/TGO solar occultations

**M. Luginin[1], A. Trokhimovskiy[1], A. Fedorova[1], D. Belyaev[1], N. Ignatiev[1], O. Korablev[1], F. Montmessin[2], and A. Grigoriev[3]**

[1] Space Research Institute (IKI), Moscow, Russia.

[2] LATMOS, UVSQ Université Paris-Saclay, Sorbonne Université, CNRS, Paris, France.

[3] Australian National University (ANU), Canberra, Australia.

Corresponding author: Mikhail Luginin (mikhail.luginin@cosmos.ru)

**Abstract**

Mesospheric $CO_2$ clouds are one of two types of carbon dioxide clouds known on Mars. We present observations of mesospheric $CO_2$ clouds made by Atmospheric Chemistry Suit (ACS) onboard the ESA-Roscosmos ExoMars Trace Gas Orbiter (TGO). We analyzed 1663 solar occultation sessions of Thermal InfraRed (TIRVIM) and Middle InfraRed (MIR) channels of ACS covering more than two Martian years that contain spectra of 2.7 μm carbon dioxide ice absorption band. That allowed us to unambiguously discriminate carbon dioxide ice aerosols from mineral dust and water ice aerosols, not relying on the information of atmospheric thermal conditions. $CO_2$ clouds were detected in eleven solar occultation observations at altitudes from 39 km to 90 km. In five cases, there were two or three layers of $CO_2$ clouds that were vertically separated by 5-15 km gaps. Effective radius of $CO_2$ aerosol particles is in the range of $0.1-2.2$ μm. Spectra produced by the smallest particles indicate a need for a better resolved $CO_2$ ice refractive index. Nadir optical depth of $CO_2$ clouds is in the range $5\times10^{-4}-4\times10^{-2}$ at both 2.7 μm and 0.8 μm. Asymmetrical diurnal distribution of detections observed by ACS is potentially due to local time variations of temperature induced by thermal tides. Two out of five cases of carbon dioxide cloud detections made by the TIRVIM instrument reveal the simultaneous presence of $CO_2$ ice and $H_2O$ ice aerosols. Temperature profiles measured by the Near InfraRed (NIR) channel of ACS are used to calculate $CO_2$ saturation ratio $S$ at locations of carbon dioxide clouds. Supersaturation $S > 1$ is detected in only 5 out of 19 cases of $CO_2$ cloud layers; extremely low values of $S < 0.1$ are found in 9 out of 19 cases.

**Keywords**: Atmospheres, structure; Infrared observations; Mars, atmosphere; Occultations; Spectroscopy.

**Key Points:**

- Detection of $CO_2$ mesospheric clouds relies on observing the 2.7 μm absorption band

- $CO_2$ clouds were detected in 11 solar occultation observations at 39-90 km altitudes
- Effective radius of $CO_2$ aerosol particles is in range of 0.1−2.2 μm
- Two cloud layers were formed by $CO_2$ and $H_2O$ ice aerosols
- $CO_2$ cloud layers with S>1 detected in 5/19 cases, S<0.1 found in 9/19 cases

**1. Introduction**

Carbon dioxide clouds are detected on Mars in the polar night troposphere and in the mesosphere in different seasons (Clancy et al., 2017). They are called polar and mesospheric $CO_2$ clouds. The latter were first identified by Clancy & Sandor (1998) based on the Pathfinder descent images showing presence of blue detached aerosol layers above 70 km (Smith et al., 1997) and observed temperatures below the $CO_2$ condensation level (Schofield et al., 1997).

The first long-term mapping of mesospheric $CO_2$ clouds covering Martian years (MY) 24-26 (we use numbering of MY proposed by Clancy et al. [2000]) at local time 13-14 h was made by Thermal Emission Spectrometer (TES) and Mars Orbiter Camera (MOC) onboard Mars Global Surveyor (MGS) (Clancy et al., 2007). They were detected at 60–80 km at equator (15°N– 15°S) mostly at solar longitudes ($L_S$) 20–40° and 140–160°. Spatial and temporal distributions of observed mesospheric $CO_2$ clouds suggested influence of thermal tides that was supported by climate modeling (González-Galindo et al., 2011; Määttänen et al., 2010; Montmessin et al., 2007), although additional cooling by gravity waves appears to be necessary to reach the temperatures required for $CO_2$ condensation (Spiga et al., 2012; Yiğit et al., 2015). Spatially-resolved nadir images of these clouds obtained by the Thermal Emission Imaging System (THEMIS) instrument onboard the Mars Odyssey spacecraft showed the same altitudinal, seasonal and spatial distribution (McConnochie et al., 2010).

The first unambiguous identification of $CO_2$ aerosols in mesospheric clouds that did not require temperature information was made by the Observatoire pour la Minéralogie, l'Eau, les Glaces et l'Activité (OMEGA) near infrared imaging spectrometer onboard the Mars Express spacecraft (Montmessin et al., 2007). In the presence of carbon dioxide clouds, OMEGA spectra demonstrated one or two distinct peaks inside a saturated $CO_2$ gaseous absorption band centered at 4.3 μm. Similar spectral pattern was also observed by the Planetary Fourier Spectrometer (PFS) onboard Mars Express (Aoki et al., 2018). Another spectroscopic identification of mesospheric $CO_2$ clouds was made from the Compact Reconnaissance Imaging Spectrometer for Mars (CRISM) data onboard the Mars Reconnaissance Orbiter (MRO) limb spectra in 0.4–4 μm range (Vincendon et al., 2011; Clancy et al., 2019).

During the night, mesospheric $CO_2$ clouds were detected from stellar occultation observations in the UV by the SPectroscopy for the Investigation of the Characteristics of the Atmosphere of Mars (SPICAM UV) instrument onboard Mars Express (Montmessin et al., 2006) and by the Imaging UltraViolet Spectrograph (IUVS) onboard the Mars Atmosphere and Volatile EvolutioN (MAVEN) spacecraft (Jiang et al., 2019). They were located at altitudes 80–110 km. In both works identification of carbon dioxide aerosol was based on simultaneously retrieved temperature profiles.

Recently, analysis of solar occultation observations performed by Nadir and Occultation for MArs Discovery (NOMAD) instrument onboard the Trace Gas Orbiter (TGO) resulted in discovery of $CO_2$ clouds at morning and evening terminators of Mars (Liuzzi et al., 2021). It was based both on spectroscopic and temperature identifications of aerosol type. In most cases, their seasonal and spatial distribution is consistent with previous observations of mesospheric $CO_2$ clouds detected at dayside.

In this work, we present observations of mesospheric $CO_2$ clouds based on solar occultation measurements by the Atmospheric Chemistry Suite (ACS) onboard TGO. To distinguish $CO_2$ aerosol particles from dust and water ice particles, we study spectra in the 2.7−2.8 μm region that contains a characteristic carbon dioxide absorption band. This is the first time that the described aerosol spectral feature of $CO_2$ ice is directly observed on Mars. In addition, measured in parallel ACS temperature profiles are used to calculate $CO_2$ saturation ratio within locations of carbon dioxide clouds, which is crucial for understanding condensation processes on Mars and modeling of clouds' formation.

## 2. Data set

TGO is a part of the ExoMars joint mission between the European Space Agency (ESA) and the Russian space agency (Roscosmos). ACS is a set of three spectrometers (Near InfraRed [NIR], Middle InfraRed [MIR], and Thermal InfraRed [TIRVIM]) with a primary goal to study trace gasses present in the Martian atmosphere and provide long-term monitoring of the temperature and atmospheric composition (Korablev et al., 2018). Previously, ACS solar occultation observations of aerosols revealed properties of water ice and dust particles during the 2018 Global Dust Storm (Luginin et al., 2020; Stcherbinine et al., 2020) and over two MY (Stcherbinine et al., 2022), discovered supersaturation of water vapor in presence of clouds and above them (Fedorova et al., 2020, 2023), and presented evidence of heterogeneous uptake of HCl on water ice particles (Luginin et al., 2024).

### 2.1. TIRVIM instrument

The TIRVIM channel is a Fourier-transform spectrometer (Korablev et al., 2018; Shakun et al., 2018) covering spectral range of 1.7–17 μm and working both in nadir and solar occultation geometry. We use measurements in a so-called "climatology" mode which is characterized by relatively good signal-to-noise ratio (SNR) and low measurement time (0.4 s) at the expense of lower spectral resolution and coarser spectral sampling ($\simeq 1$ cm$^{-1}$ and $\simeq 0.5$ cm$^{-1}$ respectively).

We use spectral points obtained within the 1,760–5,200 cm$^{-1}$ (2–6 μm) spectral range. Twenty wavenumbers outside of strong gas absorption bands, including those at ~3 μm, are used for general characterization of aerosol microphysical properties and separation of water ice particles from mineral dust and carbon dioxide ice aerosols: 1,760, 2,510, 2,600, 2,700, 2,800, 2,900, 3,027, 3,072, 3,263, 3,400, 3,446, 3,919, 4,033, 4,116, 4,346, 4,478, 4,584, 4,646, 5,030, and 5,200 cm$^{-1}$ (Set 1), following the approach from Luginin et al. (2020). To distinguish between dust and $CO_2$ aerosol particles, we analyze 2.7−2.8 μm $CO_2$ absorption feature which includes two aerosol absorption peaks: combination modes $2v_2 + v_3$ and $v_1 + v_3$ located at

~3600 cm$^{-1}$ and ~3708 cm$^{-1}$ respectively. Spectral widths (FWHM) of both peaks are ~1 cm$^{-1}$ (Isokoski et al., 2013). For CO$_2$ ice detection and retrieval of particles' properties, we use Set 1, all spectral points that fall within the 8 cm$^{-1}$ windows centered at CO$_2$ absorption peaks (Set 2) as well as those in 3550−3750 cm$^{-1}$ range not included in Set 2 and taken with 10 cm$^{-1}$ sampling (Set 3). A description of spectral points used in the study is given in Table 1.

To increase the SNR, all spectral points are averaged using a moving average over a 2-km altitude window. After this operation, resulting TIRVIM effective field of view (FOV) equals ~12 km. SNR of Set 2 equals ~100. Additionally, Set 1 is averaged over a 50-cm$^{-1}$ spectral window, providing ≥ 10$^3$ SNR. Spectral points from Set 3 are averaged over a 10-cm$^{-1}$ spectral window, which gives ~300 SNR.

**Table 1.** Description of spectral points used in the study.

| Instrument | Spectral points | | |
|---|---|---|---|
| TIRVIM | Set 1 | Set 2 | Set 3 |
| | 1,760, 2,510, 2,600, 2,700, 2,800, 2,900, 3,027, 3,072, 3,263, 3,400, 3,446, 3,919, 4,033, 4,116, 4,346, 4,478, 4,584, 4,646, 5,030, and 5,200 cm$^{-1}$. | Spectral points in 3596−3604 and 3704−3712 windows with ≃0.5 cm$^{-1}$ sampling. | Spectral points in 3550−3750 cm$^{-1}$ windows with 10 cm$^{-1}$ sampling excluding Set 2 windows. |
| MIR | Spectral points in 3704−3712 window with ≃0.05 cm$^{-1}$ sampling. | Spectral points in 3550−3750 cm$^{-1}$ windows with 2 cm$^{-1}$ sampling excluding the 3704−3712 window. | |
| NIR | 6351 cm$^{-1}$    7258 cm$^{-1}$ | 10126 cm$^{-1}$ | 11682 cm$^{-1}$    13109 cm$^{-1}$ |

**2.2. MIR instrument**

The MIR channel is a cross-dispersion echelle spectrometer dedicated to solar occultation measurements in the 2.2–4.4 μm range with a resolving power ~30 000 (Korablev et al., 2018; Trokhimovskiy et al., 2015). The diffraction orders are dispersed by a primary echelle grating to a secondary, steerable diffraction grating, resulting in a fully used 2−D detector array. In this work, we focus on secondary grating position #4 and diffraction orders 215–224, covering wavenumber range 3599–3774 cm$^{-1}$.

Every diffraction order is recorded on a detector as a stripe consisting of several rows. The number of rows depends on the wavelength range and equals 31−39 for diffraction orders

224−215 respectively. In this work, for each diffraction order we use a single row corresponding to the center of the stripe. The MIR effective FOV, with account for the integration time, equals ~ 1 km.

Combination modes of solid $CO_2$ $2\nu_2 + \nu_3$ (3600 cm$^{-1}$) and $\nu_1 + \nu_3$ (3708 cm$^{-1}$) are measured in diffraction orders 215 and 221 respectfully. We don't use ~⅙ from each side of each diffraction order due to imperfections of calibrations ('bending' of the continuum near the edges of the detector, e.g. see Figure 2 in Stcherbinine et al., 2020). As a consequence, combination mode $2\nu_2 + \nu_3$, which is located at the edge of diffraction order 215, is disregarded during the MIR data analysis. In the spectral range 3704−3712 cm$^{-1}$, we use all MIR spectral points with ~0.05 cm$^{-1}$ sampling; outside of this window, spectral points are taken with 2 cm$^{-1}$ sampling. We do not apply additional averaging due to much higher MIR SNR ($> 10^3$).

### 2.3. NIR instrument

The NIR channel is an echelle spectrometer with a selection of diffraction orders by an acousto-optical tunable filter (Korablev et al., 2018). ACS NIR covers the 0.7–1.7 μm spectral range in diffraction orders 101 through 49 with a resolving power of $\lambda/\Delta\lambda \approx 28{,}000$. During an occultation, NIR conducts measurements in 10 preselected diffraction orders in 2 s, allowing to retrieve information on gaseous abundance and aerosol microphysics. The NIR effective FOV is determined by the slit width (1.2 arcmin) and exposure time and is equal to ~1 km.

The approach on using the NIR data to characterize aerosols is the same as in Luginin et al. (2020): the NIR spectral region does not contain strong aerosol absorption bands, therefore NIR data can not be used for aerosol type determination and serves as a supplement to TIRVIM or MIR data for better definition of aerosol microphysical properties. We use "aerosol" orders 78 (10,052–10,170 cm$^{-1}$) and 90 (11,500–11,734 cm$^{-1}$) without strong gas absorption bands, as well as orders 49 (6318–6387 cm$^{-1}$), 56 (7217-7300 cm$^{-1}$), and 101 (13,016–13,170 cm$^{-1}$), which contain $CO_2$, $H_2O$, CO, and $O_2$ gaseous absorption bands. From the "gaseous" orders, slant aerosol transmission is obtained as a byproduct of gaseous abundance retrieval (Fedorova et al., 2020, 2023). SNR for NIR solar occultation measurements varied from 600 to 3000 depending on the averaging of detector lines.

During an occultation, the effective FOV of TIRVIM is several times larger than that of the NIR or MIR. When coupling NIR data with TIRVIM data, we apply a moving average over a 10-km window to NIR spectra profiles to simulate a coarser FOV of TIRVIM. This procedure additionally improves the SNR by a factor of ~2. When analyzing MIR and NIR data, such procedure is not needed.

### 2.4. Observations

TGO started its science operations in the spring 2018. As of this writing, MIR and NIR channels are performing nominally, while TIRVIM has stopped its scientific program due to its Stirling cryocooler end of life on 2nd December 2019.

In this work, we analyze 910 TIRVIM "climatology" solar occultation observations measured from 28th April 2018 ($L_S$ 166.5° of MY 34) to 2 December 2019 ($L_S$ 115.1° of MY 35), and 864 MIR solar occultation observations with secondary grating position #4 received from 23 April 2018 ($L_S$ 164.0° of MY 34) to 10 April 2022 ($L_S$ 206.4° of MY 36). NIR channel worked together with TIRVIM (MIR) in 826 (791) observations. For the sake of brevity, hereafter we will write "TIRVIM data" instead of "TIRVIM and NIR data" and "MIR data" instead of "MIR and NIR data". During 111 observations, TIRVIM and MIR instruments operated simultaneously. Totally, in this work we analyze 1663 different solar occultation observations that are shown in Figure 1.

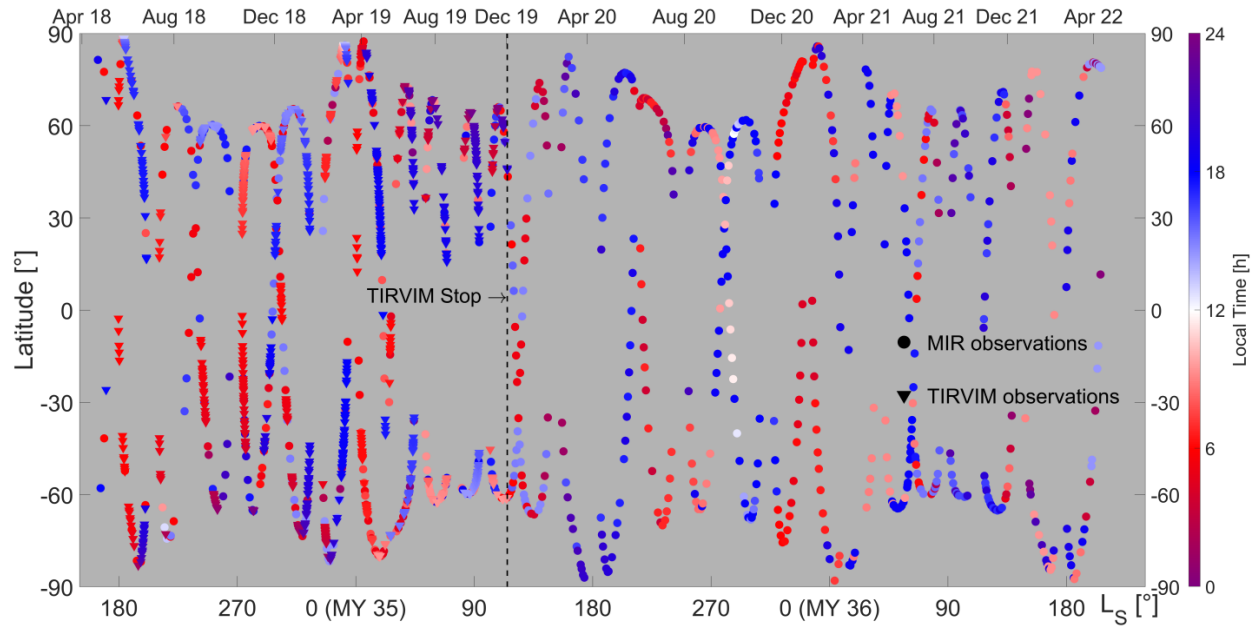

**Figure 1**. Temporal and latitudinal distribution of solar occultation observations analyzed in the work. Circles and triangles correspond to MIR and TIRVIM observations, respectively. Color code represents local time of observation. Vertical dashed line marks the 2nd of December 2019 when TIRVIM has stopped its scientific operation.

### 3. Data processing

#### 3.1. Aerosol extinction retrieval

The algorithm for retrieval of aerosol extinction mostly follows the procedure described in Luginin et al. (2020). Inside the atmosphere at altitude $z$, the solar radiance $I(z, \lambda)$ is attenuated by aerosols and gasses while passing along the line-of-sight (LOS). In this work, all altitudes are taken above the areoid. The ratio of $I(z, \lambda)$ to the reference spectrum $I_0(\lambda)$ recorded outside of the atmosphere defines a spectrum of atmospheric transmission at altitude $H$: $Tr(z, \lambda) = \frac{I(z,\lambda)}{I_0(\lambda)}$. Since the majority of spectral points are located outside of strong gaseous absorption bands, the main difference from our previous work is the processing of transmission in the region 3550−3750 cm$^{-1}$ (2.7−2.8 μm) which is reach on strong $CO_2$ gaseous lines.

Influence of gaseous lines in the transmission spectra is accounted for as follows. First, we make a correction to the wavenumber values using positions of gaseous absorption lines. Next, we calculate forward transmission model of $CO_2$ gas for a number of atmospheric layers that are defined by the geometry of each solar occultation observation. Gaseous absorption cross-sections are calculated line-by-line with the use of spectral line parameters from HITRAN 2016 database and pressure and temperature profiles retrieved from NIR (Fedorova et al., 2023) or

MIR (Belyaev et al., 2022) channels. $CO_2$ number density is calculated assuming 95.5% volume mixing ratio. Modeled gaseous transmission at different altitudes $Tr_g(z, \lambda)$ is convoluted with gaussian instrument function with FWHM of 0.16 cm$^{-1}$ for MIR and 0.94 cm$^{-1}$ for TIRVIM. Figure 2 illustrates $Tr_g$ in the 3550−3750 cm$^{-1}$ region using the transmission from the egress occultation #5401 at 33, 50 and 70 km; imaginary part of $CO_2$ ice refractive index from Warren (1986) indicates position of carbon dioxide ice absorption bands. Finally, aerosol transmission is calculated as $Tr_a(z, \lambda) = \frac{Tr(z,\lambda)}{Tr_g(z,\lambda)}$. The uncertainty of aerosol transmission $Tr_a$ in the region 2.7−2.8 μm is estimated as $\delta_{Tr_a} = \sqrt{\left(\frac{\delta_{Tr}}{Tr_g}\right)^2 + S_{Tr}^2}$, where $\delta_{Tr}$ is the uncertainty of transmission $Tr$, and $S_{Tr}^2$ is the sample variance of the transmission.

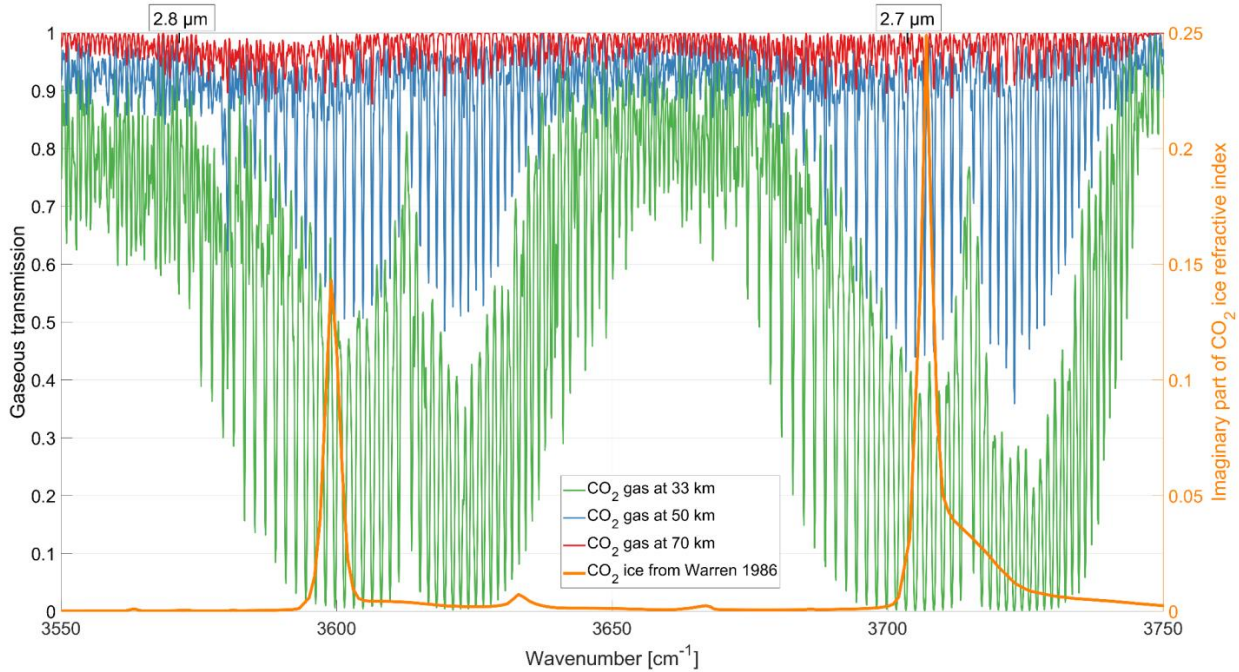

**Figure 2.** Left axis: Modeled $CO_2$ gaseous transmission spectrum from the egress occultation #5401 at 33 km (green), 50 km (blue) and 70 km (red). Right axis: imaginary part of $CO_2$ ice refractive index from Warren (1986), orange.

Examples of three typical transmission spectra patterns near the 3708 cm$^{−1}$ absorption band from MIR data are shown in Figure 3. Relatively large particles ($r_{\text{eff}}$ = 1−2 μm) produce a very deep and broad minimum ($Tr_a \simeq$ 0−0.1, $\Delta\nu \simeq$ 2−3 cm$^{-1}$) which is followed by a maximum ($Tr_a \simeq$ 0.4); transmission outside of the absorption peak is asymmetric, $Tr_a$ at shorter wavenumbers is lower (Figure 3, panel a). Smaller particles ($r_{\text{eff}}$ = 0.2−0.5 μm) produce a symmetrical absorption line with FWHM ~ 1 cm$^{-1}$ and $Tr_a \simeq$ 0.4-0.6 in the center of the CO2 feature and $Tr_a \simeq$ 0.90-0.95 outside of it (Figure 3, panel b). The smallest particles ($r_{\text{eff}} \leq$ 0.1 μm) produce a similar pattern with lower transmission values: $Tr_a \simeq$ 0.8-0.9 in the minimum and unity outside of the absorption line (Figure 3, panel c).

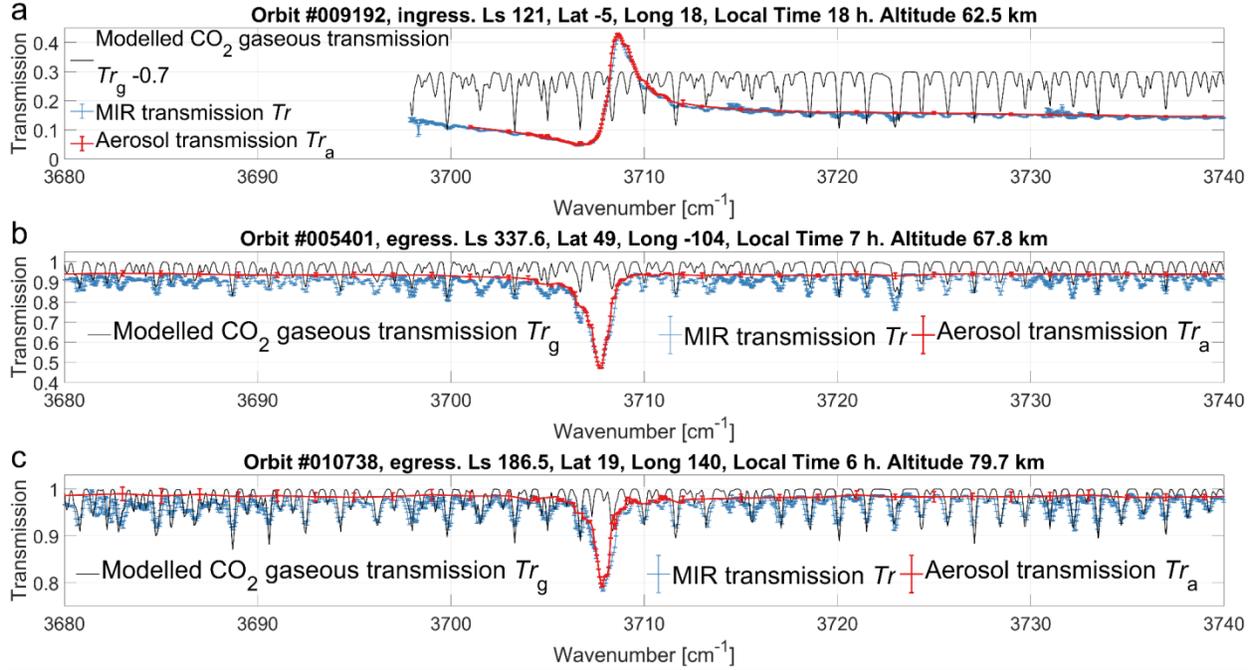

**Figure 3**. Examples of MIR transmission spectra in the 3680−3740 cm$^{-1}$ spectral region for different CO$_2$ ice particles. Measured transmission $Tr$ containing both aerosol and gaseous contribution is shown in blue, modeled CO$_2$ gaseous transmission $Tr_g$ is shown in black, retrieved aerosol transmission $Tr_a$ is shown in red. a) Case of relatively large particles ($r_{eff}$ = 1−2 μm). Note that in this panel 0.7 offset was subtracted from $Tr_g$ for clarity. b) Case of particles with $r_{eff}$ = 0.2−0.5 μm. c) Case of the smallest particles ($r_{eff} \leq$ 0.1 μm). Error bars correspond to 1−σ uncertainty level.

The slant optical depth $\tau$ with associated uncertainties $\delta_\tau$ is deduced from the Beer-Lambert law as $\tau(z,\lambda) = -\ln\ln\left(Tr_a(z,\lambda)\right), \delta_\tau = \frac{\delta_{Tr_a}}{Tr_a}$. Aerosol extinction $k$ with associated uncertainties $\delta_k$ is retrieved based on the standard "onion peeling" method (Rodgers, 2000), as described in Luginin et al. (2020).

### 3.2. Direct modeling of extinction

The aerosol extinction is modeled according to the classical Lorenz-Mie theory. We apply the code by Mishchenko et al. (1999) (https://www.giss.nasa.gov/staff/mmishchenko/Lorenz-Mie.html) and obtain extinction cross-section $\sigma$ produced by the ensemble of particles defined by their size distribution $n(r)$ and refractive index $m$ as $\sigma(\lambda, m) = \int_0^\infty \pi r^2 Q(\lambda, r, m) n(r) dr$, where $Q$ is the efficiency factor of extinction of a particle with size $r$ and refractive index $m$ at a wavelength $\lambda$.

For dust aerosols, we adopt refractive indices from Wolff et al. (2009). For water ice aerosols, we use refractive indices from Warren & Brandt (2008) and Clapp et al. (1995) measured at different temperatures and apply a temperature-wavenumber interpolation scheme described in Luginin et al. (2020), Table 1 in Section 3.3). As input temperature vertical profiles, we adopt data retrieved from the simultaneous NIR (Fedorova et al., 2023) or MIR (Belyaev et al., 2022) observations, or predictions from the Mars Climate Database v5.3 (Millour et al., 2018) if neither is available.

As seen from the high-resolution (0.1 cm$^{-1}$) solid-state IR absorbance spectra of pure $CO_2$ ice (Isokoski et al., 2013), optical properties of carbon dioxide ice depend on the temperature in the 2.7 μm region. When moving from 15 K to 75 K, FWHM of the combination modes decreases by ~1 cm$^{-1}$. For carbon dioxide ice, data on refractive indices retrieved with high spectral resolution ($\Delta v < 1$ cm$^{-1}$) and for different temperatures are quite scarce. For the 2.7 μm band of crystalline $CO_2$, only one paper by Gerakines & Hudson (2020) contains optical constants retrieved with $\Delta v = 0.2$ cm$^{-1}$ at 70 K in the spectral region 500−4000 cm$^{-1}$. For modeling of $CO_2$ aerosols, we adopt refractive indices from Gerakines & Hudson (2020) in the spectral region 3580−3720 cm$^{-1}$ and from Warren (1986) at other wavenumbers. Data from Warren (1986) is a compilation of different sources; in the 2.7 μm band, it is based on the data from Wood & Roux (1982) that was measured with $\Delta v = 4$ cm$^{-1}$ at 80 K. Figure 4 shows adopted real and imaginary parts of refractive indices of dust from Wolff et al. (2009), of water ice from Warren & Brandt (2008), and of carbon dioxide ice from Gerakines & Hudson (2020) and Warren (1986).

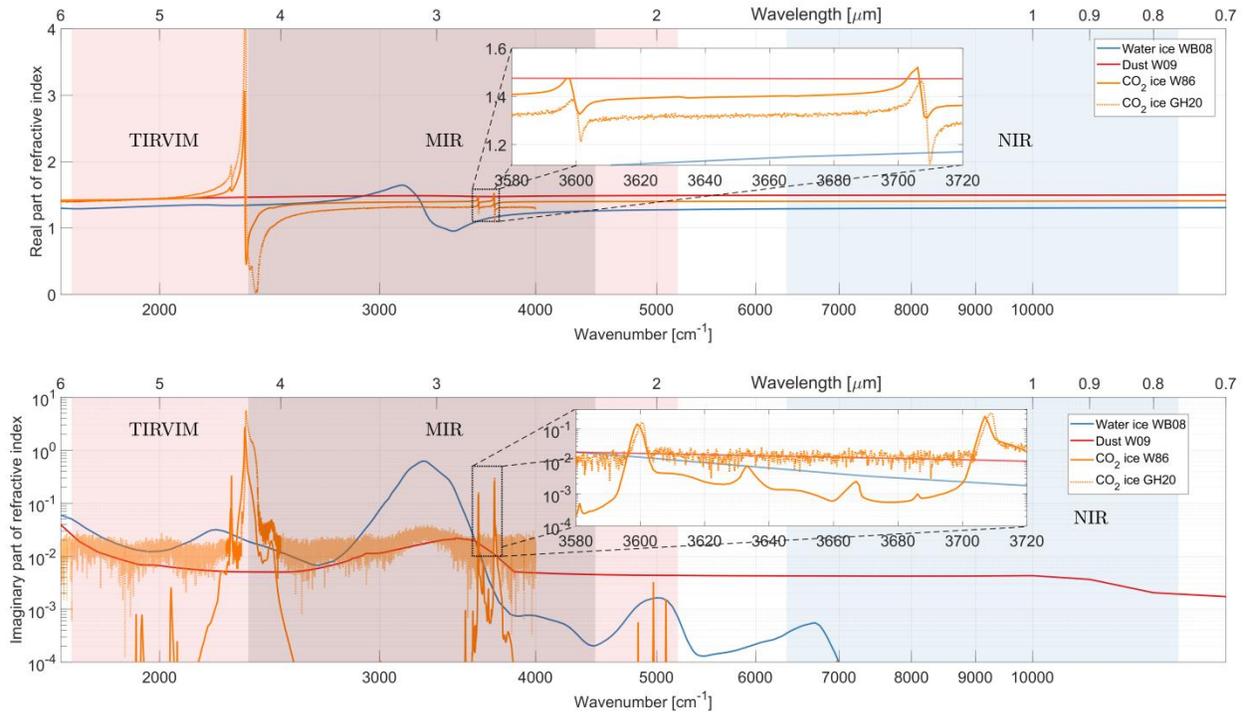

**Figure 4**. Adopted real (top panel) and imaginary (bottom panel) parts of refractive indices of dust (red lines) from Wolff et al. (2009), of water ice (blue lines) from Warren & Brandt (2008), and $CO_2$ ice from Warren (1986) and from Gerakines & Hudson (2020) (orange solid and dotted lines respectively). Inlet shows spectral region containing $CO_2$ ice combination modes $2v_2 + v_3$ at 3600 cm$^{-1}$ and $v_1 + v_3$ at 3708 cm$^{-1}$. Spectral ranges of TIRVIM, MIR and NIR are marked as red, gray and blue areas, respectively.

To describe the particle size distribution of aerosols, we use log-normal distribution $n(r) = \frac{1}{\sqrt{2\pi}\sigma_g r} exp\left(-\frac{(lnr - lnr_g)^2}{2ln^2\sigma_g}\right)$, whose parameters $r_g$ and $\sigma_g$ are directly connected to effective radius $r_{eff}$ and effective variance $v_{eff}$ as $r_{eff} = r_g exp(2.5 ln^2 \sigma_g), v_{eff} = exp(ln^2 \sigma_g) - 1$.

We have computed look-up tables of $\sigma$ for ensembles of particles made from mineral dust, water ice and $CO_2$ ice for $r_{eff}$ in the range 0.1–20 μm with 0.01 μm step and for $v_{eff}$ in the range 0.1–1.0 with 0.1 step plus 0.05 value at all wavelengths chosen for aerosol retrievals. For water ice, the variation with temperature is also accounted for, with values distributed in the range 100–270 K with 5 K step. Extinction cross sections of ensemble of $CO_2$ particles for $r_{eff} =$ 0.1−2 μm and $v_{eff} = 0.1$ in the 3580−3720 cm$^{-1}$ spectral region is shown in Figure 5. Aerosol extinction coefficient $k$ is readily calculated from $\sigma$ and the total number of particles per unit volume $N$ (number density) as $k(\lambda) = N\sigma(\lambda)$.

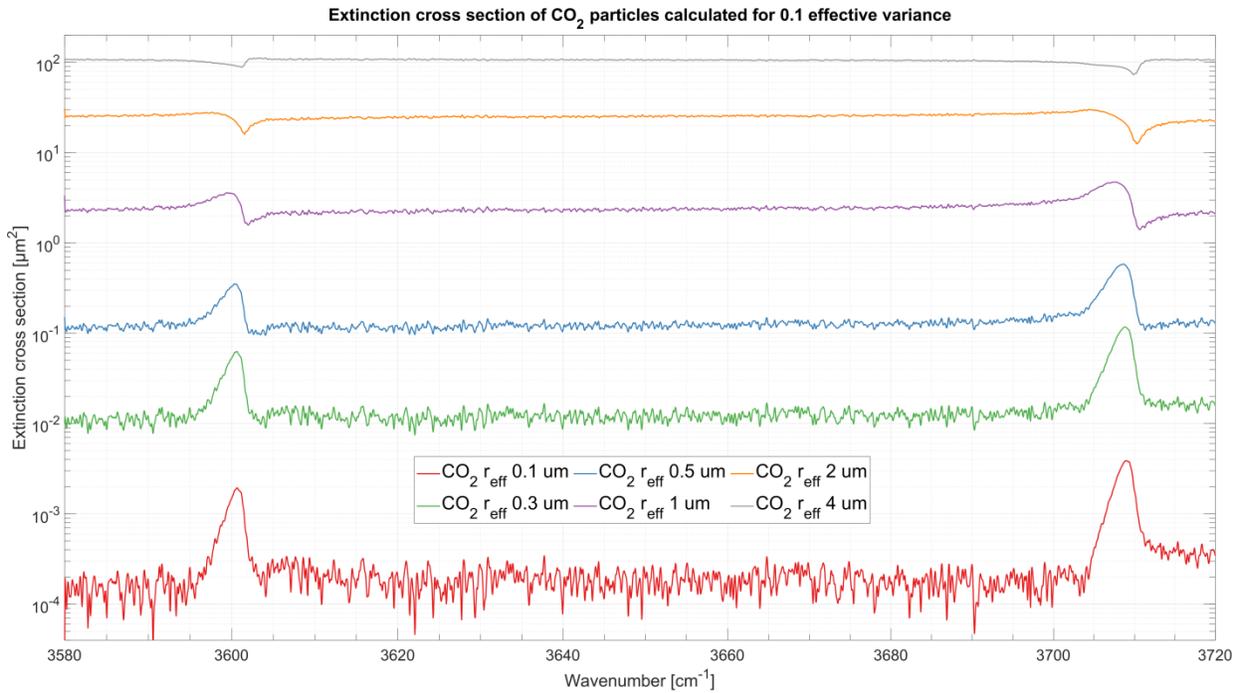

**Figure 5**. Extinction cross section of $CO_2$ particles with log-normal distribution calculated for $v_{eff} = 0.1$ and different values of $r_{eff}$ : 0.1 μm in red, 0.3 μm in green, 0.5 μm in blue, 1.0 μm in pink, 2.0 μm in orange, and 4.0 μm in grey. Refractive index in this spectral range is taken from Gerakines & Hudson (2020).

### 3.3. Retrieval of aerosol microphysical properties

The procedure for aerosol properties retrieval is substantially the same as the one described in Luginin et al. (2020). Measurements at different altitudes are processed independently. To retrieve effective radius, number density, and mass loading profiles, we solve an optimization problem by finding a minimum of merit function $\chi^2 = \frac{1}{M-p}\sum_{i=1}^{M}\frac{\left(\varkappa_i - \varkappa_i^{mod}\right)^2}{\delta_{\varkappa_i}^2}$,

where the subscript $i$ denotes wavelength from 1 to $M$, $\varkappa_i = \frac{k(\lambda_i)}{k(\lambda_0)}$ and $\varkappa_i^{mod} = \frac{k^{mod}(\lambda_i)}{k^{mod}(\lambda_0)} = \frac{\sigma^{mod}(\lambda_i)}{\sigma^{mod}(\lambda_0)}$ are experimental and modeled normalized extinction coefficients, $p$ is the number of free parameters. Wavelength $\lambda_0$ is chosen independently in each individual case and corresponds to the wavelength with the lowest uncertainties to minimize the uncertainty of normalized experimental extinction $\delta_\varkappa$, which is calculated using the propagation of error relationship $\delta_{\varkappa_i} = \varkappa_i \cdot \sqrt{\left(\frac{\delta_k(\lambda_i)}{k(\lambda_i)}\right)^2 + \left(\frac{\delta_k(\lambda_0)}{k(\lambda_0)}\right)^2}$.

Composition of aerosols is *a priori* unknown, therefore we use six competing models in our fitting procedure: an unimodal distribution consisting of either dust, water ice or $CO_2$ ice or simultaneous presence of two out of three types of aerosols, forming a bimodal log-normal distribution. In the latter case, extinction coefficient is composed of two components $k(\lambda) = N_1\sigma_1(\lambda) + N_2\sigma_2(\lambda) = N_1(\sigma_1(\lambda) + \gamma_2\sigma_2(\lambda))$, $\varkappa_i = \frac{\sigma_1(\lambda_i) + \gamma_2\sigma_2(\lambda_i)}{\sigma_1(\lambda_0) + \gamma_2\sigma_2(\lambda_0)}$, where $\sigma_i(\lambda)$, $i = 1,2$ is defined in Section 3.2, and $\gamma = \frac{N_2}{N_1}$ is the number densities ratio of two modes.

To find a minimum of the merit function, we use a simple exhaustive search through the tabulated values. We take $r_{eff}$ and $v_{eff}$ as free parameters in the unimodal distribution case and $r_{eff,1}$, $v_{eff,1}$, $r_{eff,2}$, $v_{eff,2}$, and $\gamma$ in bimodal distribution case, $\gamma$ is varied in the range $10^{-4}-10^4$.

For each of the six competing models, the fit with the minimal value $\chi^2_{j,\,min}$ is found, subscript $j$ designates the model number. The model with the lowest $\chi^2_{j,\,min}$ is considered the best, unless several models provide $\chi^2_{j,\,min} < 1$. In that case, this observation is disregarded.

During the retrieval process, it was determined that a wavenumber shift to the extinction model was necessary to fit the aerosol data. It was true for both instruments, TIRVIM and MIR. The value of the shift was always towards the shorter wavenumbers. It changed from one occultation to another and equaled 0.5–2.0 cm$^{-1}$ for MIR and 1.0–1.2 cm$^{-1}$ for TIRVIM. In addition, small (0.2–0.4 cm$^{-1}$) variations in the shift value within the same occultation were observed from the MIR data. Shifts in the observed $CO_2$ ice spectra could indicate variations in temperature. Quirico and Schmitt (1997) observed a 4 cm$^{-1}$ shift in position of carbon dioxide ice absorption bands toward shorter wavenumbers when moving from 21 K to 180 K. Our observed shifts suggest temperatures greater than 70 K of Gerakines & Hudson (2020), and are qualitatively in agreement with observed temperatures.

Example fits for cases of $CO_2$ ice particles from MIR and TIRVIM are shown in Figure 6. Panels a and b of Figure 6 contain two measurements of relative extinction spectrum near the 3708 cm$^{-1}$ peak recorded during the egress MIR occultation #15965. They are separated by only 1.5 km and illustrate a very steep transition from ~1 μm particles to ~0.1 μm particles. Panel c of Figure 6 shows relative extinction spectrum in the region 2500-4700 cm$^{-1}$ with zoom in on 3600 cm$^{-1}$ and 3708 cm$^{-1}$ peaks produced by 0.8 μm particles that was recorded during the ingress TIRVIM occultation #6452.

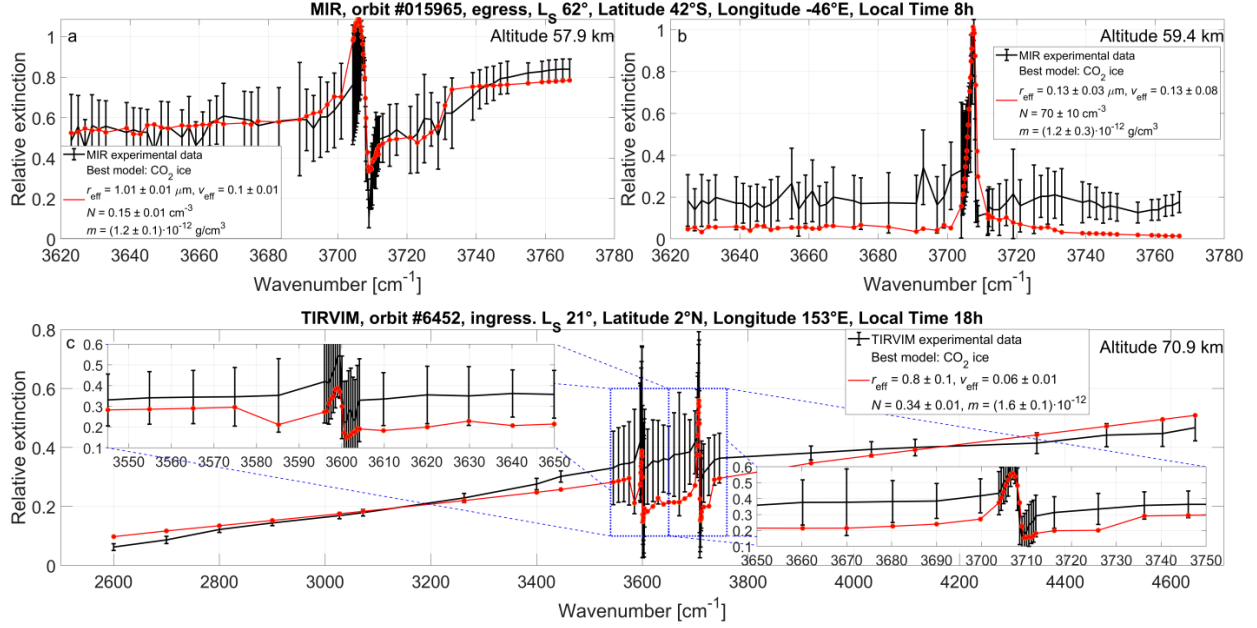

**Figure 6**. Example fits illustrating different observations of $CO_2$ ice particles. Relative extinction $\varkappa$ obtained from the experiment is shown in black, best fitting curve is shown in red. (a) MIR egress observation #15965 at 57.9 km, $r_{eff}$ = 1.0 μm. (b) The same observation as in (a) but at 59.4 km, $r_{eff}$ = 0.1 μm. (c) TIRVIM ingress observation #6452 at 70.9 km, $r_{eff}$ = 0.8 μm. Error bars of the experimental data correspond to 1−σ uncertainty level.

Once a satisfying fit is found, and all parameters are retrieved, the corresponding number density is calculated as $N = \frac{1}{M}\sum_{i=1}^{M}\frac{k(\lambda_i)}{\sigma(\lambda_i)}$. In the bimodal case, number densities of mode 1 and mode 2 are calculated as $N$ and $\gamma N$, respectively. Mass loading of the mode $j$ is calculated as $m_j = \frac{4}{3}\pi N_j \rho_j \int_0^\infty r^3 n(r, r_{eff,j}, v_{eff,j}) dr$, where ρ is assumed to be 2.5 g cm$^{-3}$ for dust particles, 0.9 g cm$^{-3}$ for water ice particles, and 1.7 g cm$^{-3}$ for $CO_2$ ice particles.

Uncertainties of the retrieved free parameters ($\delta_{r_{eff,j}}, \delta_{v_{eff,j}}, \delta_\gamma$) are calculated from the diagonal elements of the covariance matrix. Uncertainty of the number density for the unimodal case $\delta_N$ and for mode 1 $\delta_{N_1}$ in the bimodal case is estimated as $\delta_N^2 = \frac{1}{M-1}\sum_{i=1}^{M}\frac{\delta_k^2(\lambda_i)}{\sigma^2(\lambda_i) + \frac{1}{M-1}\sum_{i=1}^{M}\left(N - \frac{k(\lambda_i)}{\sigma(\lambda_i)}\right)^2}$. In the bimodal case, to get uncertainty of mode 2 number density $\delta_{N_2}$, $\delta_\gamma^2 N^2$ should be added to the previous equation. Uncertainty of mass loading is estimated as $\delta_{m_j} = m_j\sqrt{9\left(\frac{\delta_{r_{eff,j}}}{r_{eff,j}}\right)^2 + \left(\frac{\delta_{N_j}}{N_j}\right)^2}$.

### 3.4. Saturation ratio of $CO_2$

The saturation vapor pressure of $CO_2$ over solid has been measured by Giauque & Egan (1937) and recalculated from their data in form of Antoine's equation by the National Institute of Standards and Technology (NIST[1]) as follows

$$P_{sat}(T) = 6.81228 - \frac{1301.679}{T - 3.494},$$

where $P$ is the pressure in bars, and $T$ is the temperature in kelvin. Azreg-Aïnou (2005) has shown that this formula is valid from 195 K down to 65 K, which is well below the minimal temperatures in the atmosphere of Mars.

The saturation ratio of $CO_2$ $S$ is defined as $S = \frac{P_{CO_2}}{P_{sat}}$. The uncertainties of $S$ are estimated from the error propagation relation using the uncertainties of the retrieved pressure and temperature.

For calculation of $CO_2$ saturation ratio, we use pressure and temperature profiles retrieved from ACS NIR and/or ACS MIR channel (Belyaev et al., 2022; Fedorova et al., 2023). The comparison and validation of NIR and MIR temperature profiles has been done in a number of works (Alday et al., 2019, 2021; Belyaev et al., 2021, 2022; Fedorova et al., 2020, 2023).

### 4 Results and Discussion

### 4.1. Simultaneous detection of $CO_2$ clouds by TIRVIM and MIR

Out of 111 simultaneous TIRVIM and MIR solar occultation observations, only one resulted in detection of $CO_2$ clouds, namely egress occultation #5401 (L$_S$ = 338° of MY 34). In Figure 7, we compare vertical profiles of aerosol extinction coefficient $k$ at different wavenumbers, mass loading $m$, effective radius $r_{eff}$, and number density $N$ of aerosols retrieved independently from TIRVIM and MIR data during this occultation.

---

[1] https://webbook.nist.gov/cgi/cbook.cgi?ID=C124389&Mask=4

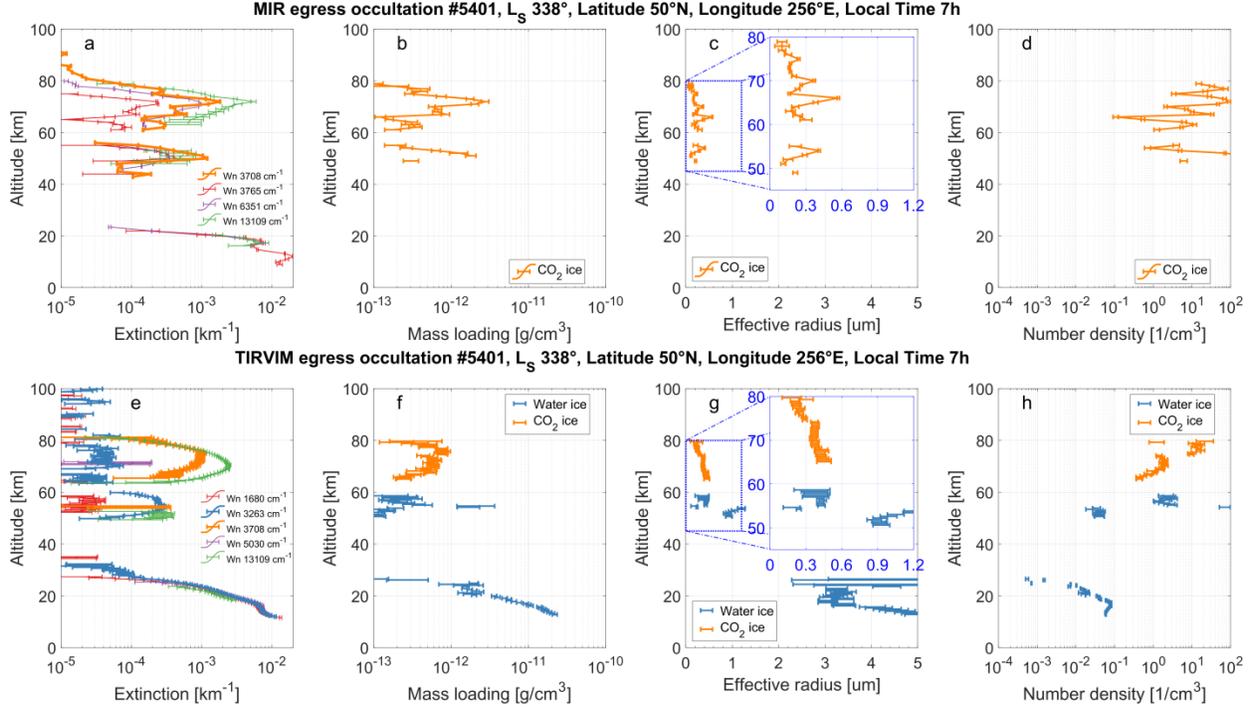

**Figure 7.** Detection of $CO_2$ ice clouds at egress occultation #5401 from the MIR (top panels) and from the TIRVIM (bottom panels) data. Vertical profiles of aerosol extinction coefficient (panels a, e), mass loading (panels b, f), effective radius (panels c, g), and number density (panels d, h). Extinction inside the aerosol carbon dioxide ice (water ice) absorption band as well as corresponding to carbon dioxide ice (water ice) retrieved products are shown in orange (blue). Error bars of the experimental data correspond to 1−σ uncertainty level.

TIRVIM, due to its extremely wide spectral range, which includes $H_2O$ ice and $CO_2$ ice absorption bands, can unambiguously distinguish all three types of aerosols on Mars; on the other hand, MIR (in the position #4) can confidently detect $CO_2$ aerosols with much higher accuracy due to its superior SNR. Another big difference between the instruments is their vertical resolution: ~12 km for TIRVIM against ~1 km for MIR. As a result, MIR discovers a lot more details in the aerosol vertical profiles. Finally, borders of the cloud layers retrieved by different instruments are slightly shifted due to different acquisition times.

The aerosol structure at the egress occultation #5401 can be divided in three layers: the main one that is formed by water ice particles and extends up to at least 27 km and two detached layers that are located at 49−55 km and 61−80 km. A clear gap in ~5 km between the two detached layers is detected by both instruments.

The higher detached layer is composed of $CO_2$ ice particles. It is vertically not uniform and in turn can be divided in four sublayers as revealed by MIR (Panels a-d of Figure 7). Inside of each sublayer, effective radius and number density, and, as a consequence, mass loading and extinction show variations around the local maximum. Effective radius lies in 0.1−0.6 μm range with a median value of 0.2 μm (Panel c of Figure 7). Nadir optical depth $\tau_{nad}(\lambda) = \int_{H_1}^{H_2} k(\lambda, z) dz$ of the whole detached layer at 3708 cm$^{-1}$ (2.7 μm, MIR) and 13109 cm$^{-1}$ (0.8 μm, NIR) equals, respectively, $1.7 \cdot 10^{-2}$ and $2.5 \cdot 10^{-2}$. A list of all $CO_2$ detection cases with corresponding observation parameters and retrieved aerosol properties is gathered in Table 2.

From the TIRVIM extinction data, the higher detached layer spans from 63 to 80 km, although aerosol extinction coefficient at 3708 cm$^{-1}$ is detected only starting from 65 km (Panel e of Figure 7). At altitudes 63−65 km, both dust and $CO_2$ ice unimodal distribution models provide $\chi^2 < 1$, and, according to our retrieval procedure, are disregarded. Two sublayers inside the higher detached layer can be extracted from the TIRVIM data. Effective radius is in the 0.2−0.5 µm range with median value of 0.4 µm. Optical depth of the whole detached layer at 3708 cm$^{-1}$ (TIRVIM) equals $10^{-2}$.

The lower detached layer is composed of both $H_2O$ ice and $CO_2$ ice particles. The presence of two sublayers of carbon dioxide aerosols is revealed by MIR, the lower one being faint and barely detectable (Panels a and c of Figure 7). Effective radius of $CO_2$ ice particles in the lower detached layer is in 0.1−0.4 µm range with a median value of 0.2 µm. In case of TIRVIM, SNR of extinction coefficient at 3708 cm$^{-1}$ in this detached layer is slightly above unity, and, as a consequence, $CO_2$ ice particles are only detected but their properties are not retrieved. Instead, TIRVIM detects $H_2O$ ice particles with effective radius 0.3−1.2 µm at 51−59 km. Optical depth of the detached layer at 3708 cm$^{-1}$ (MIR) and 13109 cm$^{-1}$ (NIR) equals, respectively, $4\cdot10^{-3}$ and $3\cdot10^{-3}$.

**Table 2.** Summary of TIRVIM and MIR occultations with detected $CO_2$ clouds and retrieved microphysical properties of the $CO_2$ ice particles. Columns with $r_{eff}$, $N$ and $M$ contain minimal-median-maximal values. $S_{max}$ contains maximal value of saturation ratio within the extended by 5 km altitude boundaries calculated from NIR and MIR pressure and temperature data, respectively. Column with $\tau$ shows nadir optical depth of the $CO_2$ cloud layer at 3708 cm$^{-1}$ and 13109 cm$^{-1}$. $H_2O$ ice column indicates simultaneous detections (✓) of $CO_2$ ice and $H_2O$ ice particles from TIRVIM data, possible detections (?) from MIR data, and no detections (✗) of $H_2O$ ice particles.

| Occ. | Channel | $L_s$ [°] (MY) | Lat. | Eastern Long. | Loc. time [h] | $z$ [km] | $r_{eff}$ [μm] | $N$ [cm$^{-3}$] | $M$ [10$^{-14}$ g/cm$^3$] | $S_{max}$ NIR; $S_{max}$ MIR | $\tau \times 10^{-3}$ | $H_2O$ ice |
|---|---|---|---|---|---|---|---|---|---|---|---|---|
| 3502_E | TIRVIM | 244.4 (34) | 27°S | −65° | 5.2 | 72–82 | 0.6–0.7–0.8 | 0.02–0.1–0.3 | 5–20–30 | 0.08; − | 3; 4 | ✓ |
| 5401_E | MIR | 337.6 (34) | 50°N | −104° | 6.8 | 49–55 | 0.1–0.2–0.4 | 0.7–6–200 | 20–60–200 | [7; 9]·10$^{-4}$ | 4; 3 | ? |
|  |  |  |  |  |  | 61–79 | 0.1–0.2–0.6 | 0.1–10–90 | 7–40–300 | 0.07, 0.03 | 17; 25 | ✗ |
|  | TIRVIM |  |  |  |  | 47–56 | —[1] | —[1] | —[1] | —[1] |  | ✓ |
|  |  |  |  |  |  | 65–80 | 0.2–0.4–0.5 | 0.4–2–20 | 20–50–100 | 0.07; 0.03 | 10; 22 | ✗ |
| 6452_I | TIRVIM | 20.7 (35) | 1°S | 152° | 18.0 | 59–78 | 0.2–0.8–0.9 | 0.03–0.3–5 | 6–80–200 | 10; − | 18; 21 | ✗ |
| 6602_E | TIRVIM | 26.4 (35) | 6°S | 7° | 6.1 | 76–86 | 0.1–0.15–0.21 | 10–40–500 | 20–100–200 | 1; − | 7; 9 | ✗ |
| 9192_I | MIR | 121.0 (35) | 5°S | 19° | 17.9 | 55, 57[2] | 2.2 ± 0.1, 1.01 ± 0.05[2] | 0.03, 0.4[2] | 200, 200[2] | 0.03; 0.02 | 4; 3 | ? |
|  |  |  |  |  |  | 62–70 | 0.1–0.8–1.9 | 0.1–2–600 | 70–500–1000 | 0.8; 0.02 | 40; 37 | ? |
| 10738_E | MIR | 186.5 (35) | 21°N | 140° | 6.1 | 50–65 | 0.1–0.1–0.3 | 3–20–60 | 20–30–40 | [8; 3]·10$^{-3}$ | 3; 6 | ✗ |
|  |  |  |  |  |  | 72–74 | 0.2–0.25–0.3 | 0.2–0.2–0.6 | 2–2–3 | 0.2; 0.06 | [5; 8]·10$^{-2}$ | ✗ |
|  |  |  |  |  |  | 79–90 | 0.1–0.1–0.3 | 0.3–30–70 | 2–10–40 | 1; 0.3 | 3; 0.6 | ✗ |
| 14059_E | MIR | 349.1 (35) | 30°S | −91° | 5.8 | 45–62 | 0.13–0.16–0.4 | 0.4–10–30 | 6–20–60 | [2; 2]·10$^{-3}$ | 7; 20 | ✗ |
| 14501_I | MIR | 7.3 (36) | 12°N | 13° | 18.1 | 48–52 | 0.1–0.2–0.3 | 5–13–70 | 30–50–60 | [5; 5]·10$^{-3}$ | 1; 3 | ? |
|  |  |  |  |  |  | 58–65 | 0.1–0.11–0.23 | 5–40–100 | 8–40–50 | 0.2; 0.3 | 2; 1 | ✗ |
|  |  |  |  |  |  | 71–79 | 0.1–0.16–0.3 | 0.3–4–20 | 3–5–10 | 4; 4 | 0.5; 0.8 | ✗ |
| 15965_E | MIR | 62.1 (36) | 42°S | −46° | 7.4 | 53–59 | 0.29–1–1.1 | 0.02–0.2–70 | 15–25–80 | 0.1; 0.07 | 3; 3 | ✗ |

| 16825_E | MIR | 92.9 (36) | 54°S | 143° | 8.7 | 58−63 | 0.1−1.9−2.2 | 0.1−0.3−70 | 4−500−2000 | 0.3; 20 | 29; 31 | ✗ |
|---|---|---|---|---|---|---|---|---|---|---|---|---|
| 19076_E | MIR | 184.1 (36) | 49°N | −23° | 6.1 | 39−54 | 0.1[3] | 40−50−60[3] | 20−30−40[3] | [3; 4]·10[-4] | 5; 3 | ✗ |
| | | | | | | 69−76 | 0.1[3] | 30−70−100[3] | 20−40−60[3] | 0.05; 0.09 | 5; 4 | ✗ |
| | | | | | | 80−84 | 0.1[3] | 16−20−40 | 10−13−20 | 1; 0.3 | 1; 0.5 | ✗ |

[1] Properties of $CO_2$ clouds were not retrieved from TIRVIM data at these altitudes due to the lower SNR compared to MIR. Instead, TIRVIM detected presence of $H_2O$ particles.

[2] Two MIR measurements were made in this $CO_2$ cloud layer. Here for $r_{eff}$, $N$ and $M$, we give exactly the values at probed altitudes instead of minimal-median-maximal values throughout the layer.

[3] All best fits in these cases give the same $r_{eff} = 0.1$ μm, see Section 4.2.

### 4.2. Detection statistics of CO$_2$ clouds and their microphysical properties

All occultation sessions described in the 2.4 Section have been processed. Figure 8 shows a map of all cases of CO$_2$ clouds detections from the TIRVIM and MIR solar occultation using the 2.7 μm band. CO$_2$ ice clouds were detected in 4 out of 910 TIRVIM sessions (0.4 % occurrence) and in 8 out of 864 MIR sessions (0.9 % occurrence). Lower frequency of CO$_2$ clouds detection from the TIRVIM data could be related to its much lower intrinsic SNR (before averaging). As was written earlier, both TIRVIM and MIR channels operated and detected CO$_2$ clouds during the egress occultation #5401. Overall, we have 11 sessions with CO$_2$ clouds detections out of 1663 combined ACS solar occultation observations (0.7 % occurrence). This value of occurrence is close to 1.0 % (4 out of 412) from SPICAV UV (Montmessin et al., 2006) and 0.6 % (2 out of 309) from IUVS (Jiang et al., 2019) observations. Liuzzi et al. (2021) reports 26 detections covering period L$_S$ = 340° of MY 34 to L$_S$ = 325° of MY 35. Though authors did not detail the total number of analyzed sessions, according to [nomad.aeronomie.be](nomad.aeronomie.be) site, there were 4575 NOMAD solar occultations in this period. The estimate of NOMAD occurrences is therefore 0.6 % and is close to our result.

Analysis of the diurnal distribution shows a significant asymmetry between the terminators: only three detections of the CO$_2$ clouds were made at the evening one while eight happened at the morning one. One of the possible explanations is the local time variation of temperature induced by the thermal tides. Fedorova et al. (2023) studied the differences in temperature between the morning and the evening terminators at different altitudes, latitudes and seasons from the ACS NIR solar occultation data. Despite the complicated latitude-altitude dependence of the temperature difference, that was also observed by other instruments (e.g. Kleinböhl et al., 2013; Fan et al., 2022; Guerlet et al., 2022), temperature was generally higher at the evening terminator in the low and middle latitudes at altitudes 40-80 km, where the most of the detected CO$_2$ clouds is observed.

CO$_2$ clouds were detected from 39 km up to 90 km. While in principle solar occultations can sound altitudes close to the surface, it is impossible with our method to detect CO$_2$ clouds at altitudes ≲30 km due to saturation of CO$_2$ gaseous absorption (see Figure 2). In five out of eleven cases, there were two or three layers of CO$_2$ clouds that were separated by aerosol-free gaps, with a total number of 19 CO$_2$ ice layers. A typical gap width was 5 km, reaching 15 km in a single case. Altitude range of CO$_2$ cloud detections is wider than 60−80 km from the TES observations (Clancy et al., 2007) and 55−75 km range from the CRISM data (Clancy et al., 2019), not as high as 80−105 km from the SPICAM UV measurements (Montmessin et al., 2006) and 80−100 km from the UIVS observations (Jiang et al., 2019), and almost identical to 40−95 km from the NOMAD solar occultations (Liuzzi et al., 2021). In addition, in 2 out of 26 NOMAD cases there were two CO$_2$ ice layers separated by ~5 km gaps.

Values of retrieved effective radius of CO$_2$ particles are in range 0.1−2.2 μm with a median value of 0.6 μm. We must note that 0.1 μm value falls on the edge of the parameter search space (0.1−20 μm). In this case, the quality of fit was reduced: the FWHM of the absorption peak from the model was equal to 2.5 cm$^{-1}$ (both for $r_{\text{eff}}$ = 0.01 μm and for $r_{\text{eff}}$ = 0.1 μm with $v_{\text{eff}}$ = 0.1) compared to 1.3 cm$^{-1}$ from the MIR data. This fact indicates that the model poorly represents extinction cross section for very small CO$_2$ ice particles due to a low spectral resolution of refractive index, and that the actual effective radius of the particle could be <0.1 μm. A high-resolution (0.1 cm$^{-1}$) absorption spectra of solid-state CO$_2$ ice gives 1.1 cm$^{-1}$

FWHM of 3708 cm$^{-1}$ band at 75 K (Isokoski et al., 2013) which is close to our experimental data. Hence, a better resolved $CO_2$ ice refractive index (better than 0.1 cm$^{-1}$ and, preferably, for a wide range of temperatures) is needed for a proper modeling of small $CO_2$ particles and retrieval of their microphysical properties from the 2.7 μm band. Our results are in a good agreement with most previous observations: $r_{eff}$ = 0.1−0.3 μm from the Pathfinder lander images (Clancy & Sandor, 1998), $r_{eff} \leq 1.5$ μm from the TES data (Clancy et al., 2007), $r_{eff}$ = 1–3 μm with median values of 2.0–2.3 μm from the OMEGA observations (Määttänen et al., 2010), $r_{eff}$ = 0.5–2 μm and $r_{eff}$ = 0.3−2.2 μm from the CRISM measurements (Vincendon et al., 2011 and Clancy et al., 2019, respectively), $r_{eff}$ = 0.08–0.13 μm from the SPICAM UV data (Montmessin et al., 2006), and $r_{eff}$ = 0.09–0.11 μm from the UIVS observations (Jiang et al., 2019). From the NOMAD data, Liuzzi et al. (2021) has reported retrievals of effective radii from 0.1 up to 4.5 μm; such large values exceed both previous studies and our observations.

Retrieved mass loading was distributed in range $2\times10^{-14}$–$2\times10^{-11}$ g/cm$^3$ with a median value of $2\times10^{-13}$ g/cm$^3$. Nadir optical depth of the $CO_2$ clouds at 3708 cm$^{-1}$ (2.7 μm) and 13109 cm$^{-1}$ (0.8 μm) was in the range $5\times10^{-4}$–$4\times10^{-2}$ with median values $5\times10^{-3}$ at both wavenumbers. Montmessin et al. (2006) retrieved comparable $\tau_{nad}$ = $6\times10^{-3}$–$5\times10^{-2}$ at 0.2 μm from the SPICAM UV stellar occultations, while $\tau_{nad}$ = 0.01–0.5 at 1 μm from the OMEGA observations (Määttänen et al., 2010) and typical value $\tau_{nad}$ = 0.2 at 0.5 μm from the CRISM measurements (Vincendon et al., 2011) were larger than our results, as expected from the nadir observations.

An extremely thin cloud was once detected on the ingress occultation #9192 when only two MIR measurements at 55 km and 57 km were made within the $CO_2$ cloud layer. It was composed of quite large particles with $r_{eff}$ = 2.2 ± 0.1 μm and $r_{eff}$ = 1.01 ± 0.05 μm, $M = 2\times10^{-12}$ g/cm$^3$ at both altitudes and extremely small nadir optical depth of $5\times10^{-5}$ and $8\times10^{-5}$ at 3708 cm$^{-1}$ and 13109 cm$^{-1}$.

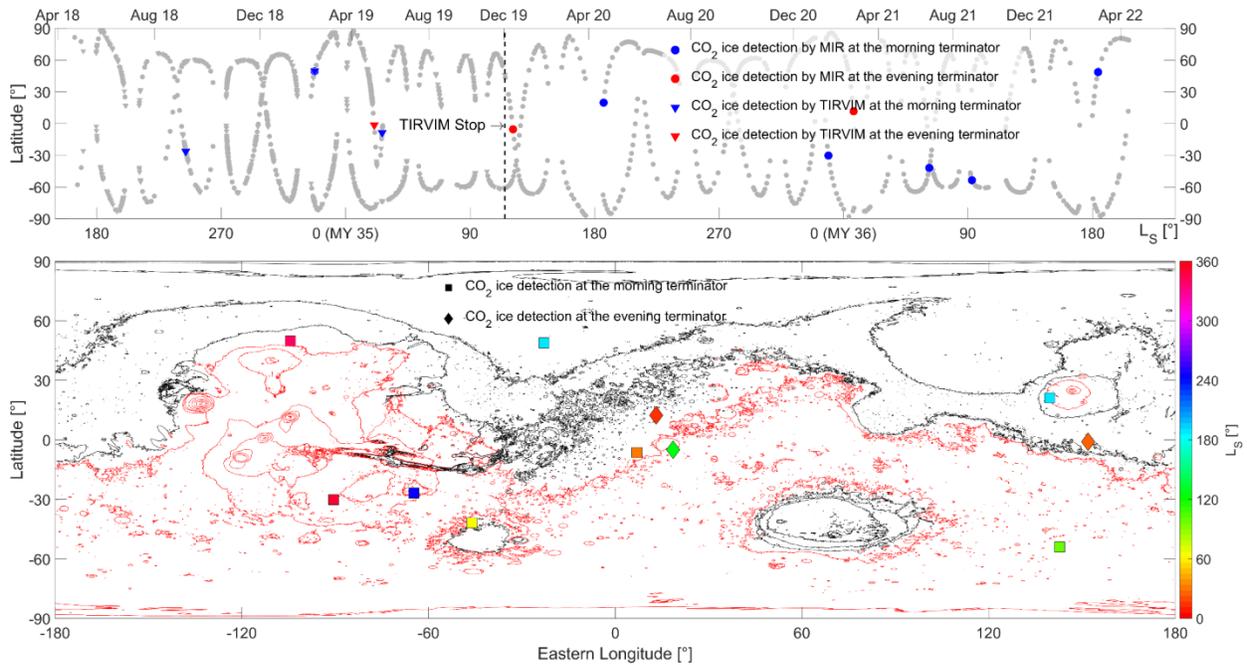

**Figure 8.** Map of the $CO_2$ clouds detections from the 2.7 μm absorption band from the ACS instrument. Top: Seasonal and latitudinal distribution of solar occultation observations. Circles

and triangles correspond to MIR and TIRVIM observations. No $CO_2$ clouds detection is shown in grey, detection at the morning (evening) terminator is in blue (red). Bottom: Latitudinal-longitudinal map of the $CO_2$ clouds detections from the ACS instrument with the solar longitude given by the color code. Squares and triangles represent detections made at the morning and at the evening terminators, respectively. Contour lines represent the Mars Orbiter Laser Altimeter (MOLA) altimetry data. Black lines correspond to negative values taken with 1 km step, red lines — positive values with 3 km step.

### 4.3. $CO_2$ clouds and saturation ratio

In our work, all aerosol observations are accompanied by measurements of atmospheric pressure and temperature from the same solar occultation session, which are used for calculations of saturation ratio $S$. Surprisingly, supersaturation $S > 1$ was detected in only 5 out of 19 cases of $CO_2$ cloud layers (26 %); extremely low values $S < 0.1$ were found in 9 out of 19 cases (47 %) (see Table 2). Figure 9 shows two examples of $T$, $P$ and $S$ profiles retrieved independently from NIR and MIR data. Similarity between the two instruments is evident. The former example (#5401, egress), as was discussed above, featured two $CO_2$ clouds layers, both having $S < 0.1$. The latter one, ingress occultation #14501, had three layers with $CO_2$ particles, and only at the altitudes of the topmost one $S > 1$ was found. Such behavior when only one out three $CO_2$ layers had $S > 1$ was observed in three solar occultation observations.

Discovery of $CO_2$ clouds existing at extremely low values of $S$, even with a supersaturated layer at highest altitudes, is unexpected, and requires a comprehensive microphysical modeling coupled with a GCM to account for evolution of temperature profiles for interpretation of this phenomenon. A simple explanation, that particles in the supersaturated region are condensing, growing, and gravitationally falling in the subsaturated region where they are detected, is quantitatively inadequate. For example, in case of #14501 occultation (bottom of Figure 9), an estimation of carbon dioxide ice particles lifetime using measured $P$ and $T$ profiles (following the "linearized model" from Listowski et al., 2013) results in $\ll 1$ s timescale at altitudes <70 km where above freezing temperatures were observed. It is several orders of magnitudes lower compared to the time needed for a particle at a sedimentation speed (calculated using expression 20.4 from Jacobson, 2005) to reach the altitude of the lowest $CO_2$ layer, which means that a particle under the observed conditions should sublimate long before reaching those altitudes.

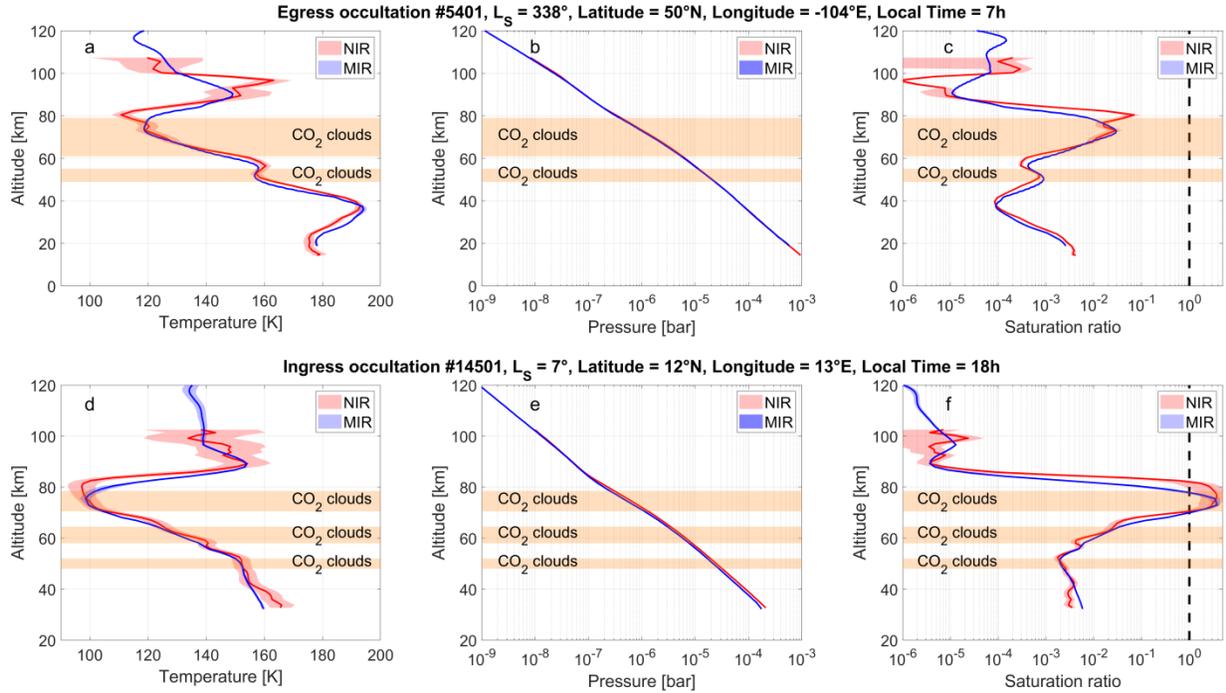

**Figure 9.** Examples of pressure (panels a, d), temperature (panels b, e) and saturation ratio (panels c, f) profiles retrieved from NIR (red) and MIR (blue) data from egress occultation #5401 (top) and ingress occultation #14501 (bottom). Bold lines and filled areas correspond to best fits and 1−σ uncertainty, respectively. Altitudes where $CO_2$ clouds were detected are marked by horizontal orange bars.

Among the 1663 unique ACS solar occultation observations analyzed in this work, $S > 1$ was observed in 28 measurements (< 2 %). As was pointed out before, $CO_2$ clouds were detected only in five of them, while in other cases there were no aerosol layers at the altitudes of saturation ratio maximum. A similar pattern was observed by Jiang et al. (2019) from the IUVS data: out of seven occultations of τ-Sco star that detected supersaturation only one featured aerosol layer. These results raise the questions about the source and microphysical properties of condensation nuclei in the Martian mesosphere.

Thus, we confirm that observed $S > 1$ condition alone is not enough to make a confident conclusion about $CO_2$ clouds presence and may lead to a substantial overestimation of $CO_2$ clouds occurrence. On the other hand, from our observations it follows that the $S > 1$ condition coinciding with an aerosol layer always corresponds to a $CO_2$ cloud detection and can be used as a proxy for an unambiguous $CO_2$ identification, but will lead to numerous overlooked $CO_2$ clouds.

### 4.4. Departure from spherical symmetry

Solar and stellar occultation observations are usually interpreted in an assumption of a spherically symmetric atmosphere, which implies, for aerosol studies, a combination of attenuating layers, variable with altitude and uniform in latitude and longitude. We will call this extinction model 'standard' in this Section. Panel a of Figure 9 illustrates retrieved from the ingress occultation #14501 typical extinction profiles in the 2D geometry (we show only a half of the atmosphere) with three $CO_2$ cloud layers A, B and C located at different altitudes. However, the atmosphere is not necessarily symmetric and many models may be constructed to fit the experimental data.

One of the possible alternative extinction distributions is shown in panel c of Figure 10. It was constructed to satisfy the following conditions:
1) Reproduce the observed aerosol transmission profiles $Tr_a$ at 3708 cm$^{-1}$;
2) Have spherically uniform aerosol layers everywhere except for altitudes of the A cloud;
3) Don't have the local maxima in extinction profiles that correspond to clouds B and C of standard model.

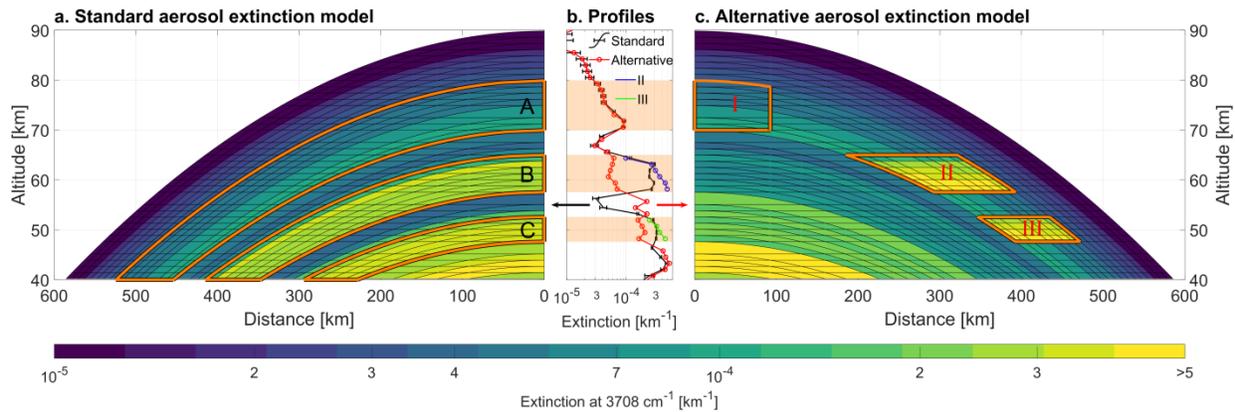

**Figure 10.** Illustration of an alternative aerosol extinction model for the ingress occultation #14501. Panel a: A standard for solar occultation spherically symmetric model. Three $CO_2$ cloud layers A, B and C (shown as orange frames) are located at different altitudes and are uniform in latitude and longitude. Panel b: vertical profiles of aerosol extinction for standard model (black), alternative model (red) and layers II (blue) and III (green) of the alternative model. Panel c: Alternative model. Three $CO_2$ cloud layers I, II, and III (shown as orange frames) are located at the same altitudes, but different latitudes/longitudes.

As a result, the alternative model has three $CO_2$ cloud layers, marked I, II, and III, that belong to the same spherical shell (are at the same altitudes), but are located at different latitudes/longitudes. The aerosol extinction of layers A and I are equal (panel b of Figure 10). Inside the layers II and III, extinction is vertically not uniform and is equal to extinction of layers B and C at higher altitudes and 2-3 times smaller at lower altitudes. The nadir optical depths of layers II and III equal $2.1 \cdot 10^{-3}$ and $1.3 \cdot 10^{-3}$, and are only marginally higher than values of layers B and C (Table 2).

Demonstrated alternative model could potentially resolve the controversy between the detected $CO_2$ clouds at lower altitudes and extremely low values of saturation that was discussed previously. Indeed, small horizontal variations of temperature (5-10 K) is enough to result in drop in saturation from ~5 to ≲1. However, it is only speculative and needs *a priori* data to justify the selection of the aerosol extinction model. Solar occultation measurements alone do not contain enough information to do so: the number of transmission measurements in a single profile is $\mathbb{N}$, while the number of sublayers is ~ $\mathbb{N}^2$. Moreover, in five out of eleven occultations $S$ was ≪ 1 for all $CO_2$ cloud layers (e.g. #5401 in Figure 9), which requires another explanation.

Also, when constructing the alternative model, we worked only with a single extinction at 3708 cm$^{-1}$ but not its spectral dependence. Considering the factors mentioned above, we conclude that while a departure from spherical symmetry is possible in some sessions, the primary conclusions and questions raised in the previous Section remain valid.

### 4.5. Simultaneous identification of H$_2$O and CO$_2$ clouds

From the TIRVIM data, two out of five cloud layers were formed by CO$_2$ and H$_2$O ice aerosols. They were located at ~50 km and ~75 km. Effective radius of water ice particles was in range 0.1−0.2 μm and smaller than CO$_2$ particles at the same altitudes in one case, while in the other 0.4−1.2 μm H$_2$O ice particles were larger compared to carbon dioxide ice.

The MIR spectra analyzed in this work does not allow an unambiguous simultaneous detection of H$_2$O and CO$_2$ aerosols due to lack of absorption signatures of mineral dust and H$_2$O ice in the studied spectral region. If both types of clouds are observed by MIR, models with CO$_2$ particles and either dust or water ice produce essentially the same quality of fit and can not be discriminated. There are four such cases (including the one when TIRVIM detected H$_2$O ice particles) out of 16 detected CO$_2$ cloud layers; they are shown as question marks in Table 2. Effective radius of potential water ice particles is in the range 0.1−0.9 μm.

Recently, Määttänen et al. (2022) has coupled a CO$_2$ cloud microphysical model (Listowski et al., 2014, 2013) with the LMD Mars Global Climate Model (Forget et al., 1999) featuring a detailed H$_2$O cloud microphysics (Navarro et al., 2014). Among other improvements, they have included a possibility for CO$_2$ ice to form on water ice crystals in addition to mineral dust as on condensation nuclei. It resulted in a significant increase in the amount of modeled mesospheric CO$_2$ ice clouds at low and middle latitudes. In our analysis, we did not study the question whether the water ice aerosols that were detected simultaneously with carbon dioxide aerosols are in the form of separate particles or encapsulated within the composite H$_2$O/CO$_2$ particles. In order to do that, a dedicated model of composite aerosol particle extinction should be added to the fitting procedure.

## 5 Conclusions

We analyzed 1663 ACS solar occultation observations that contain spectra of carbon dioxide ice combination mode $\nu_1 + \nu_3$ at 3708 cm$^{-1}$ (2.7 μm). This characteristic spectral signature directly seen in our spectra allows us to detect mesospheric CO$_2$ clouds on Mars, unambiguously discriminating them from mineral dust and water ice aerosols. Unlike other approaches that rely on the information of atmospheric thermal conditions to deduce type of the observed aerosols, we use temperature profiles from the same occultation sessions to retrieve CO$_2$ saturation ratio at the location of carbon dioxide clouds, thus providing important for future numerical modeling information on the state of the CO$_2$ atmosphere of Mars. The key results of our study are as follows:
1. Overall, CO$_2$ clouds were detected in eleven solar occultation observations. In five cases, there were two or three layers of CO$_2$ clouds that were vertically separated by gaps. Carbon dioxide ice was detected from 39 km to 90 km, typical gap width was 5 km, reaching 15 km in a single instance.

2. Effective radius of $CO_2$ aerosol particles was in the range of 0.1−2.2 μm with a median value of 0.6 μm. Spectra of $v_1 + v_3$ combination mode in case of smallest particles indicate a need for a better resolved $CO_2$ ice refractive index.
3. Mass loading ranged from $2\times10^{-14}$ to $2\times10^{-11}$ g/cm³ with a median value of $2\times10^{-13}$ g/cm³. Nadir optical depth of $CO_2$ clouds at 3708 cm$^{-1}$ and 13109 cm$^{-1}$ was in the range $5\times10^{-4}$–$4\times10^{-2}$ with median values $5\times10^{-3}$ at both wavenumbers.
4. Diurnal distribution of detections is significantly asymmetrical: only three detections at the evening terminator versus eight at the morning one. A possible explanation is local time variations of temperature induced by thermal tides.
5. Analysis of ACS TIRVIM spectra that include water ice and carbon dioxide ice absorption features revealed that two out of four cases of $CO_2$ clouds detections were accompanied by observations of $H_2O$ aerosols.
6. Supersaturation $S > 1$ was detected in only 5 out of 19 cases of $CO_2$ cloud layers; extremely low values $S < 0.1$ were found in 9 out of 19 cases. Interpretation of these observations requires a comprehensive microphysical modeling coupled with a GCM to account for evolution of temperature.

One potential extension to the current work involves investigation of the 4.3 μm carbon dioxide ice band, which may lead to detection of $CO_2$ clouds at higher altitudes due to its stronger absorption. ACS TIRVIM and ACS MIR data can be utilized for this purpose since they both include described spectra. Besides that, a systematic analysis of aerosol and temperature profiles retrieved from a comprehensive dataset of ACS NIR channel gathered during more than three Martian years could reveal climatology of $CO_2$ clouds, along with their spatial and diurnal variations.

**Declaration of competing interest**

The authors declare that they have no known competing financial interests or personal relationships that could have appeared to influence the work reported in this paper.

**Acknowledgments**

ExoMars is a space mission of ESA and Roscosmos. The ACS experiment is led by IKI, the Space Research Institute in Moscow, assisted by LATMOS in France. The work was supported by the Ministry of Science and Higher Education of the Russian government.